\documentclass[prd,aps,preprintnumbers,floats,floatfix,superscriptaddress,preprintnumbers,
showpacs,eqsecnum,nofootinbib,notitlepage,
twocolumn
]{revtex4-1}
\usepackage[utf8]{inputenc}
\usepackage{latexsym,array,theorem,mathrsfs,bm,float}
\usepackage{psfrag}
\usepackage{amsfonts,amsmath,amssymb,latexsym,array,afterpage,
theorem,mathrsfs,bm,float,epsfig,color,graphicx,tabularx,here,multirow}

\newcommand{\nn}{\nonumber \\}

\newcommand{\nablaA}{\overset{\scriptscriptstyle  \Gamma}{\nabla }}

\newcommand{\RA}{\overset{\scriptscriptstyle  \Gamma}{R }}
\newcommand{\GA}{\overset{\scriptscriptstyle  \Gamma}{G }}

\newcommand{\RAcal}{\overset{\scriptscriptstyle  \Gamma}{\mathcal{R} }}

\newcommand{\Pri}{ \Phi }
\newcommand{\Sec}{ \Psi }

\newcommand{\KA}{\overset{\scriptscriptstyle  \Gamma}{\mathcal{K} }}

\begin{document}
\baselineskip=12pt

\preprint{YITP-19-32, WU-AP/1903/19}

\title{Scalar-metric-affine theories: Can we get ghost-free theories from symmetry?
}
\author{Katsuki \sc{Aoki}}
\email{katsuki.aoki@yukawa.kyoto-u.ac.jp}
\affiliation{Center for Gravitational Physics, Yukawa Institute for Theoretical Physics, Kyoto University, 606-8502, Kyoto, Japan}
\affiliation{
Department of Physics, Waseda University,
Shinjuku, Tokyo 169-8555, Japan
}

\author{Keigo \sc{Shimada}}
\email{shimada.k.ah@m.titech.ac.jp}
\affiliation{
Department of Physics, Tokyo Institute of Technology, Tokyo 152-8551, Japan
}
\affiliation{
Department of Physics, Waseda University,
Shinjuku, Tokyo 169-8555, Japan
}

\date{\today}

\begin{abstract}
We reveal the existence of a certain hidden symmetry in general ghost-free scalar-tensor theories which can only be seen when generalizing the geometry of the spacetime from Riemannian.  For this purpose, we study scalar-tensor theories in the metric-affine (Palatini) formalism of gravity, which we call scalar-metric-affine theories for short, where the metric and the connection are independent.  We show that the projective symmetry, a local symmetry under a shift of the connection, can provide a ghost-free structure of scalar-metric-affine theories.  The ghostly sector of the second-order derivative of the scalar is absorbed into the projective gauge mode when the unitary gauge can be imposed.  Incidentally, the connection does not have the kinetic term
in these theories
and then it is just an auxiliary field.  We can thus (at least in principle) integrate the connection out and obtain a form of scalar-tensor theories in the Riemannian geometry.  The projective symmetry then hides in the ghost-free scalar-tensor theories.  As an explicit example, we show the relationship between the quadratic order scalar-metric-affine theory and the quadratic U-degenerate theory.  The explicit correspondence between the metric-affine (Palatini) formalism and the metric one could be also useful for analyzing phenomenology such as inflation.

\end{abstract}



\maketitle

\section{Introduction}
In the last decade, a great deal of attention has been paid to scalar-tensor theories including second-order derivatives of a scalar field in the Lagrangian for building models of inflation and dark energy.  The prototype theory, called the Galileon, is motivated by the decoupling limit of the Dvali-Gabadadze-Porrati model~\cite{Dvali:2000hr} where the effect of the modification of gravity is effectively described by a single scalar degree of freedom in addition to the tensor ones~\cite{Nicolis:2008in}.  
The absence of an explicit UV complete model of gravity motivates us to consider general scalar-tensor theories as an effective field theory (EFT).
The standard guidance of EFT is to include all possible terms consistent with the underlying (exact or approximate) symmetry in the Lagrangian.  However, in the context of the modified gravity, ghost-free class of scalar-tensor theories have been extensively discussed.

The second-order derivatives in the Lagrangian generally yield a fourth-order equation of motion which contains an unstable mode called the Ostrogradsky ghost.  The Galileon interactions are so special that the equation of motion becomes second-order even though the Lagrangian contains second-order derivatives.
The generalized Galileon, now dubbed as the Horndeski theory~\cite{Horndeski:1974wa,Charmousis:2011bf,Deffayet:2011gz,Kobayashi:2011nu,Ezquiaga:2016nqo}, is the most general scalar-tensor theory with the equation of motion with at most second-order derivatives.  After the discovery of the Gleyzes-Langlois-Piazza-Vernizzi (GLPV) theory~\cite{Gleyzes:2014dya,Gleyzes:2014qga}, it is recognized that the assumption of keeping the second-order equation of motion is too strong for general Ostrogradsky ghost-free theories. This led to the idea of degeneracy, in which a constraint is imposed onto the Lagrangian and eliminates the Ostrogradsky ghost. The degenerate higher order scalar-tensor (DHOST) theory is a theory where higher derivative interactions are tuned to satisfy the degeneracy conditions~\cite{Zumalacarregui:2013pma,Langlois:2015cwa,Langlois:2015skt,Crisostomi:2016czh,Achour:2016rkg,BenAchour:2016fzp}.  Furthermore, it is recently suggested that the degeneracy conditions are not necessarily satisfied in arbitrary gauge of the spacetime when assuming appropriate boundary conditions~\cite{DeFelice:2018mkq}.  The paper~\cite{DeFelice:2018mkq} proposed a more general ghost-free scalar-tensor theory, the U-degenerate theory, which satisfies the degeneracy conditions at least in the unitary gauge\footnote{In the paper~\cite{DeFelice:2018mkq}, U-degenerate theories and DHOST theories are classified: U-degenerate theories are not degenerate in an arbitrary gauge but degenerate in the unitary gauge while DHOST theories are degenerate under any gauge.  However, in the present paper, we will just call theories ``U-degenerate theories" if the Lagrangian is degenerate at least in the unitary gauge for simplicity.} in which uniform scalar field hypersurfaces correspond to constant time hypersurfaces (see also~\cite{Gao:2014soa,Gao:2018znj}).

Although a ghost mode can exist in a low energy EFT, the existence of the ghost restricts the cutoff scale of EFT to be lower than the mass of the ghost mode.  Therefore, the ghostly higher derivative terms provide just sub-leading contributions within the regime of validity of EFT.  In order that the higher derivative interactions provide leading contributions for inflation or dark energy phenomena, the interactions have to be tuned to eliminate the Ostrogradsky ghost.  In a subset of ghost-free theories, this ghost-free structure can be protected by the weakly broken Galileon symmetry~\cite{Pirtskhalava:2015nla,Santoni:2018rrx}.  On the other hand, in the generic ghost-free theories, the fine-tuned ghost-free structure seems not to be robust since no underlying symmetry apparently exist. If there are no robust structures, could they still be considered as interesting theories of EFTs for inflation/dark energy?

In the present paper, we, however, point out that the general ghost-free scalar-tensor theories indeed have a hidden symmetry which can be seen only when generalizing the geometry of spacetime.  Gravitational theories are usually formulated in Riemannian geometry, where only the metric is the independent object that characterize the intrinsic structure of space-time geometry.  The connection in Riemannian geometry is called the Levi-Civita connection which is calculated by the metric.  Meanwhile, the metric and the connection define different geometrical concepts, the inner product and the parallel transport, and thus they are independent in the first place.  Following this philosophy, one could consider a geometry with the Riemannian metric and a general affine connection: metric-affine geometry. The metric-affine geometry becomes the Riemannian geometry when the metric-compatibility condition and the torsionless condition are imposed.

It is worthwhile mentioning that both metric-affine geometry and scalars could emerge from a more fundamental perspective.  Scalars, such as dilatons, radions, axions, scalarons, and sfermions, are known famously to appear from extended theories of the standard model or certain quantum gravity models.  Whereas metric-affine geometry is expected to emerge from space-time defects~\cite{Lobo:2014nwa,Olmo:2015bha} and Riemann-Cartan geometry, which is a subset of metric-affine geometry, may come from Poincare gauge theory~\cite{Blagojevic:2003cg} or Supergravity~\cite{Freedman:2012zz}.  Thus, from such perspective scalar-tensor theories within metric-affine geometry is a field worth investigating. From here and now on, we shall call scalar-tensor theories constructed on metric-affine geometry, scalar-metric-affine theories for short. On the other hand, we will call scalar-tensor theories based on Riemannian geometry, simply scalar-tensor theories.  We will show that the projective symmetry, a local symmetry under a shift of the connection, can provide a ghost-free structure of scalar-metric-affine theories.  The ghostly part of the second-order derivative of the scalar could be taken into the gauge mode which means that the connection plays the role of a ``ghostbuster field''~\cite{Fujimori:2016udq, Fujimori:2017rcc}.  Incidentally, the connection does not have a kinetic term as long as higher curvature terms are not included and then it is just an auxiliary field.  We can thus (at least in principle) integrate the connection out and obtain a form of scalar-tensor theories~\cite{Aoki:2018lwx}.  The projective symmetry then hides in the ghost-free scalar-tensor theories.

The rest of the present paper is organized as follows.  We briefly review metric-affine geometry and the formalism of gravity based on it in Section~\ref{sec_metric-affine}.
The main result of the present paper is shown in Section~\ref{sec_minimal} where we consider scalar-metric-affine theories without non-minimal couplings to the curvature: the Lagrangian is given by the Einstein-Hilbert Lagrangian plus an arbitrary function constructed by up to the second-order covariant derivatives of a scalar field.  We will show that this class of theory is free from the Ostrogradsky ghost when the projective symmetry is imposed and the unitary gauge can be assumed.  More general scalar-metric-affine theories with non-minimal coupling to the curvature are discussed in Section~\ref{sec_non-minimal} and some ghost-free couplings are found.  We integrate the connection out and explicitly show the relation between the quadratic order scalar-metric-affine theory and the quadratic U-degenerate theory in Section~\ref{sec_quadratic}.  We make summary remarks in the last Section~\ref{summary}.  General Hamiltonian analysis of scalar-metric-affine theories is discussed in Appendix~\ref{appendix}.

\section{Metric-affine formalism of gravity}
\label{sec_metric-affine}
The intrinsic structure of the metric-affine geometry is defined in terms of two independent objects: the Riemannian metric $g_{\mu\nu}$ and the general affine connection $\Gamma^{\mu}_{\alpha\beta}$.
The covariant derivatives for a vector and a co-vector are defined by
\begin{align}
\nablaA_{\alpha}A^{\mu}
&=\partial_{\alpha}A^{\mu}+\Gamma^{\mu}_{~\beta\alpha}A^{\beta}
\,,
\\
\nablaA{}_{\alpha}A_{\mu}&=\partial_{\alpha}A_{\mu}-\Gamma^{\beta}_{~\mu\alpha}A_{\beta}\,.
\end{align}
The Riemann curvature tensor is then defined by
\begin{align}
\RA{}^{ \mu}{}_{\nu\alpha\beta}(\Gamma)&:=\partial_{\alpha}\Gamma^{\mu}_{~\nu\beta}-\partial_{\beta}\Gamma^{\mu}_{~\nu\alpha}
+\Gamma^{\mu}_{~\sigma\alpha } \Gamma^{\sigma}_{~\nu\beta }-\Gamma^{\mu}_{~\sigma \beta} \Gamma^{\sigma}_{~\nu\alpha } 
\,.
\end{align}
Since the metric and the connection are independent, in addition to the Riemann curvature tensor, there exist non-vanishing tensors characterizing the geometry;
\begin{align}
T^{\mu}{}_{\alpha\beta}&:=\Gamma^{\mu}_{~\beta\alpha}-\Gamma^{\mu}_{~\alpha\beta}
\,,\\
Q_{\mu}{}^{\alpha\beta}&:=\nablaA{}_{\mu}g^{\alpha\beta}\,,
\end{align}
which are called the torsion tensor and the non-metricity tensor, respectively.  The Riemannian geometry, which is usually used to formulate gravitational theories, is obtained by imposing two constraints, $T^{\mu}{}_{\alpha\beta}=0,Q_{\mu}{}^{\alpha\beta}=0$, under which the connection is then uniquely determined to be the Levi-Civita connection as,
\begin{align}
\Gamma^{\mu}{}_{\alpha\beta}=
\left\{ {}^{\,\, \mu}_{\alpha\beta} \right\} := \frac{1}{2}g^{\mu\nu}(\partial_{\alpha} g_{\beta\nu}+\partial_{\beta}g_{\alpha\nu}-\partial_{\nu}g_{\alpha\beta} ) \,,
\end{align}
and then the connection is no longer independent from the metric.  Thus, the only independent object in Riemannian geometry becomes the metric.

The metric-affine formalism (also called Palatini formalism) of gravity assumes that the metric and the connection are independent in the first place and gravitational theories determine dynamics of not only the metric but also the connection.  Therefore, all of the intrinsic structure of spacetime geometry is determined by gravity.  On the other hand, the metric formalism a priori assumes Riemannian geometry and gravitational theories determine the metric only; that is, gravity assumed to be essentially represented by a symmetric tensor field.

Let us first consider the metric-affine counterpart of the Einstein-Hilbert action
\begin{align}
\mathcal{L}_{\rm EH}=\frac{M_{\rm pl}^2}{2}\RA(g,\Gamma)
\,, \label{EH_action}
\end{align}
where the Ricci scalar is defined by
\begin{align}
\RA:=g^{\mu\nu}\RA{}^{\alpha}{}_{\mu\alpha\nu} \,.
\end{align}
Note that the EH action in the metric-affine formalism enjoys an additional gauge symmetry of
\begin{align}
\Gamma^{\mu}_{\alpha\beta}\rightarrow \Gamma^{\mu}_{\alpha\beta}+\delta^{\mu}_{\alpha}U_{\beta}\,, \label{gauge_inv}
\end{align}
with an arbitrary vector $U_{\beta}(x)$.  
The transformation \eqref{gauge_inv}, called the projective transformation, preserves two characteristics; the geodesic equation up to the redefinition of the affine parameter and the angle between two vectors under parallel transport~\cite{riccischouten,veblen1926projective}.  The invariance/symmetry under \eqref{gauge_inv} is dubbed the projective invariance/symmetry. There are two facts that are worth mentioning. First, projective symmetry only emerges in metric-affine geometry, and not in its subsets such as Riemann-Cartan geometry.  Secondly, the particles of the standard models are also known to inhibit projective symmetry. Some properties of the projective invariance at quantum level are discussed in \cite{Kalmykov:1994fm,Kalmykov:1994yj,Kalmykov:1995ab,Kalmykov:1998wr}.

In the case of vacuum, the equation of motion of the connection yields
\begin{align}
\Gamma^{\mu}_{~\alpha\beta}= \left\{ {}^{\,\, \mu}_{\alpha\beta} \right\} 
\,.
\end{align}
up to the gauge freedom associated with the projective symmetry.  Therefore, the connection does not introduce any new dynamical degrees of freedom and \eqref{EH_action} predicts the same results obtained by the Einstein-Hilbert action in the metric formalism~\cite{giachetta1997projective,Sotiriou:2009xt,Dadhich:2010xa,Bernal:2016lhq}.

In general, however, theories in the metric-affine formalism compute different results from their metric formalism counterpart.  To discuss general theories, we introduce the distortion tensor defined by
\begin{align}
\kappa^{\mu}{}_{\alpha\beta}:= \Gamma^{\mu}_{~\alpha\beta}- \left\{ {}^{\,\, \mu}_{\alpha\beta} \right\} 
\,,
\end{align}
which expresses the deviations from Riemannian geometry.  The curvature tensor, the torsion tensor, and the non-metricity tensor are then
\begin{align}
\RA{}^{ \mu}{}_{\nu\alpha\beta}(\Gamma)&=
R^{\mu}{}_{\nu\alpha\beta}(g)
+
2\nabla_{[\alpha}\kappa^{\mu}{}_{\nu|\beta]}
+2\kappa^{\mu}{}_{\sigma[\alpha } \kappa^{\sigma}{}_{\nu|\beta] }
\,,
\\
T^{\mu}{}_{\alpha\beta}&=2\kappa^{\mu}{}_{[\beta\alpha]} \,,
\\
Q_{\mu}^{\alpha\beta}&=2\kappa^{(\alpha\beta)}{}_{\mu}
\,,
\end{align}
where $R^{\mu}{}_{\nu\alpha\beta}$ and $\nabla_{\mu}$ are the Riemann curvature and the covariant derivatives defined by the Levi-Civita connection, respectively.  
Since only the curvature contains the derivative of $\kappa$ and it is linear, kinetic terms of $\kappa$ appear when higher curvature terms are involved in Lagrangian.  Even if such higher curvature terms exist in a full theory, one may ignore higher curvature corrections in a low energy limit.  In such case, the "dynamics" of $\kappa$ is determined by a constraint equation (at least in a low energy limit) and then can be integrated out as with the case of the Einstein-Hilbert action\footnote{In special cases, $\kappa$ is still non-dynamical even when higher curvature terms are introduced in metric-affine gravity. For example, metric-affine f(R) theories do not introduce new degrees of freedom unlike their metric formalism counterpart\cite{Sotiriou:2006qn,Sotiriou:2009xt,DeFelice:2010aj,Olmo:2011uz}.}.  In this paper, we shall focus on Lagrangian consisting of terms up to linear in the curvature so that $\kappa$ does not carry any new degrees freedom.  If the constraint yields $\kappa \neq 0$, such theory gives a different theory from its metric formalism counterpart, i.e., theory with setting $\kappa =0$. For further discussion and reviews of metric-affine gravity see for example ~\cite{Hehl:1994ue,Gronwald:1997bx,Hehl:1999sb,Blagojevic:2013xpa,Kleyn:2004yj,Iosifidis:2019jgi}.

\section{Ghost-free scalar field from projective symmetry}
\label{sec_minimal}
In this section, we assume the Einstein-Hilbert action as the gravitational sector and consider a scalar field whose Lagrangian is assumed to be constructed by up to second-order covariant derivatives,
\begin{align}
\mathcal{L}_{\phi}=\mathcal{L}_{\phi}(g,\phi,\nablaA_{\mu} \phi, \nablaA{}_{\mu}\nablaA_{\nu} \phi) \,,
\end{align}
where the covariant derivatives are
\begin{align}
\nablaA_{\mu}\phi=\partial_{\mu}\phi \,, \quad \nablaA_{\mu}\nablaA_{\nu}\phi=\partial_{\mu}\partial_{\nu}\phi-\Gamma^{\alpha}_{~\nu\mu}\partial_{\alpha}\phi\,.
\end{align}
The total Lagrangian describing the present system is, thus,
\begin{align}
\mathcal{L}(g,\Gamma,\phi)=\frac{M_{\rm pl}^2}{2} \RA(g,\Gamma)+\mathcal{L}_{\phi}(g,\phi,\nablaA_{\mu} \phi, \nablaA{}_{\mu}\nablaA_{\nu} \phi) 
\,.  \label{Ltot}
\end{align}
Since the second-order derivative of $\phi$ contains the connection and therefore $\kappa$ as
\begin{align}
\nablaA_{\mu}\nablaA_{\nu}\phi =\nabla_{\mu}\nabla_{\nu}\phi-\kappa^{\alpha}{}_{\nu\mu}\partial_{\alpha}\phi\,,
\end{align}
the existence of $\nablaA\nablaA \phi$ in the Lagrangian changes the constraint equation of $\kappa$ whose solution would be generally given by $\kappa=\kappa(g,\phi,\nabla\phi, \nabla^2 \phi)$ (see~\cite{Aoki:2018lwx} for an explicit solution of the constraint in a scalar-metric-affine theory).  Substituting it into the Lagrangian, we obtain the form of a scalar-tensor theory as
\begin{align}
\mathcal{L}(g,\phi)=\frac{M_{\rm pl}^2}{2}R+\mathcal{L}'_{\phi}(g,\phi,\nabla\phi,\nabla^2\phi) \,.
\label{Ltot2}
\end{align}

In general, \eqref{Ltot} (or the equivalent Lagrangian \eqref{Ltot2}) must have second-order time derivatives of the scalar field yielding the Ostrogradsky ghost mode.  One way to obtain Ostrogradsky ghost-free theories is that the second time derivatives are tuned to satisfy the degeneracy conditions.  We do not assume such conditions; instead, we assume the projective symmetry to absorb the ghostly time derivatives into the gauge mode.

The EH action is invariant under the projective transformation \eqref{gauge_inv}.  Let us suppose that the scalar field Lagrangian also enjoys the projective symmetry.  Since the connection appears only in the covariant derivative of the scalar field, the projective symmetry of $\mathcal{L}_{\phi}$ is realized by the invariance under 
\begin{align}
\nablaA{}_{\mu}\nablaA_{\nu} \phi \rightarrow \nablaA{}_{\mu}\nablaA{}_{\nu}\phi - U_{\mu}\partial_{\nu}\phi
\,.
\end{align}

To see the relation between this invariance and the ghost-free property, we consider the $3+1$ decomposition.
Introducing the unit normal vector $n_{\alpha}$ to 3-dimensional spacelike hypersurfaces and the projection tensor on the hypersurfaces, defined by
\begin{align}
\gamma_{\mu\nu}:=g_{\mu\nu}+n_{\mu}n_{\nu}
\,,
\end{align}
we first decompose the first derivative of $\phi$ into the temporal part and the spatial parts,
\begin{align}
A_*:=n^{\mu}A_{\mu}\,, \quad \hat{A}_{\mu}:=\gamma_{\mu}^{\nu}A_{\nu}
\end{align}
with $A_{\mu}:=\partial_{\mu}\phi$.
The second-order derivative is then
\begin{align}
\nabla_{\mu}\nabla_{\nu}\phi &= D_{\mu}\hat{A}_{\nu} -A_* K_{\mu\nu}+2n_{(\mu} (K_{\nu) \alpha} \hat{A}^{\alpha}-D_{\nu)} A_*)
\nn
&+n_{\mu}n_{\nu}(\pounds_n A_* -\hat{A}_{\alpha}a^{\alpha}) \label{second_phi}
\end{align}
where $D_{\mu}$ is the covariant derivative associated with the spatial metric $\gamma_{\mu\nu}$, $\pounds_n$ is the Lie derivative with respect to $n^{\mu}$, $a_{\mu}:=n^{\alpha}\nabla_{\alpha}n_{\mu}$ is the acceleration, $N$ is the lapse function, and $K_{\mu\nu}:=\frac{1}{2}\pounds_n \gamma_{\mu\nu}$ is the extrinsic curvature.  The existence of $\pounds_n A_*$, which contains the second time derivative of $\phi$ and the first time derivative of $N$, in the Lagrangian is a signature of the Ostrogradsky ghost.

The paper~\cite{DeFelice:2018mkq} argued that the original degeneracy conditions obtained by~\cite{Langlois:2015cwa} are too strong in order for the theory to be free from the Ostrogradsky ghost.  They suggest that it is sufficient for the ghost-freeness to satisfy the degeneracy condition at least in the unitary gauge $\phi=\phi(t)$.  Note that in the unitary gauge, $\phi$ is not a dynamical variable.  Even so, the degeneracy of $\pounds_n A_*$ yields the Ostrogradsky ghost-free theory because $A_*=\dot{\phi}(t)/N$ in the unitary gauge and then the disappearance 
of $\pounds_n A_*$ guarantees that the lapse function is still non-dynamical.

We assume that the unitary gauge can be consistently chosen in theory \eqref{Ltot}.  In the unitary gauge, the uniform $\phi$ hypersurfaces correspond to the 3-dimensional hypersurfaces and then the unit normal vector is proportional to $\partial_{\mu}\phi$.  This leads to that the projective symmetry now becomes the invariance under
\begin{align}
\nablaA_{\mu}\nablaA_{\nu}\phi \rightarrow \nablaA_{\mu}\nablaA_{\nu}\phi+A_* U_{\mu}n_{\nu}\,,
\end{align}
for an arbitrary vector $U_{\mu}(x)$.  Comparing with \eqref{second_phi}, one can find that the term with $\pounds_n A_*$ in \eqref{second_phi} is just the projective mode.  Hence, the projective symmetry of $\mathcal{L}_{\phi}$ guarantees that the scalar field Lagrangian does not have $\pounds_n A_*$\footnote{Alternatively, one can fix the projective gauge so that $\pounds_n A_*$ does not appear in $\nablaA{}_{\mu}\nablaA{}_{\nu} \phi$.}; that is, the Lagrangian is trivially U-degenerate.

As a result, the projective symmetry of $\mathcal{L}_{\phi}$ guarantees that \eqref{Ltot} has no dependence on $\pounds_n A_*$ in the unitary gauge,
\begin{align}
\mathcal{L}_{\phi}(g,\phi,\nablaA_{\mu} \phi, \nablaA{}_{\mu}\nablaA_{\nu} \phi) =\mathcal{L}_{\phi}(t,N,\gamma_{\mu\nu},K_{\mu\nu},\kappa;D_{\mu})\,.
\end{align}
Note that the diffeomorphism invariance of \eqref{Ltot} guarantees that the Lagrangian has no explicit dependence on the shift.  
Since the action \eqref{Ltot} is algebraic in terms on the distortion tensor $\kappa$, integrating out $\kappa$ does not yield $\pounds_n A_*$.  Hence, the Lagrangian may be given by the form
\begin{align}
\mathcal{L}=\mathcal{L}(t,N,\gamma_{\mu\nu},K_{\mu\nu};D_{\mu})\,,
\label{LADM}
\end{align}
after integrating out $\kappa$.  The remaining spatial diffeomorphism invariance and the fact that the corresponding Hamiltonian to \eqref{LADM} is linear in $N^i$ give 6 first class constraints which reduce 12 degrees of freedom from the phase space.  Furthermore, since \eqref{LADM} does not have $\dot{N}$, there is a primary constraint $\pi_N \approx 0$ whose time preservation yields a secondary constraint.  Since the temporal diffeomorphism invariance is broken due to the gauge fixing, the constraints $\pi_N \approx 0$ and $\dot{\pi}_N \approx 0$ are generally second class and then reduce 2 degrees of freedom.  Whether or not there exist further constraints in \eqref{LADM}, the number of constraints is sufficient to eliminate the Ostrogradsky ghost.  Therefore, we conclude that the projective invariant Lagrangian \eqref{Ltot} has at most 3 degrees of freedom and free from the Ostrogradsky ghost.  General Hamiltonian analysis is discussed in Appendix~\ref{appendix}.

\section{Non-minimal coupling to curvature}
\label{sec_non-minimal}
We then consider non-minimal couplings to the curvature tensor keeping the projective symmetry.  When a non-minimal coupling is introduced, we need to take care of the fact that the curvature tensor has the first derivative of $\kappa$.  For instance, the non-minimal coupling to the Ricci scalar,
\begin{align}
f_1 \RA \supset 2f_1 \nabla_{\alpha} \kappa^{[\alpha\beta]}{}_{\beta} \,,
\end{align}
where $f_1$ is some function, can be rewritten, by taking integration by parts, as
\begin{align*}
-2(\nabla_{\alpha}f_1) \kappa^{[\alpha\beta]}{}_{\beta} \,.
\end{align*}
Hence, even if $f_1$ does not contain $\pounds_n A_*$, $\nabla_{\alpha}f_1$ may yield it when $f_1$ contains $A_*$ and then the ghost-free characteristic of scalar-metric-affine theories with a non-minimal coupling is not manifest.

Although whether projective invariant non-minimal couplings are generally ghost-free or not is interesting question, we only discuss some particular couplings to curvature tensors which are definitely ghost-free \footnote{For simplicity, we do not consider either torsion or non-metricity couplings in this paper. Since the torsion and the non-metricity do not contain derivatives of the distortion tensor, it is expected that these couplings do not drastically change the structure of the theories and do not lead to the Ostrogradsky ghost.
}.

\subsection{Trivially U-degenerate couplings}
We find that the following non-minimal couplings
\begin{align}
f_2 \GA{}^{\mu\nu}\nablaA_{\mu}\phi \nablaA_{\nu} \phi\,, ~ 
f_3 \GA{}^{\mu\alpha\nu\beta}\nablaA_{\mu}\phi \nablaA_{\nu} \phi \nablaA_{\alpha}\nablaA_{\beta}\phi,
\label{non_minimal}
\end{align}
do not lead to the Ostrogradsky ghost, where $f_2$ and $f_3$ are projective invariant functions of $\phi,\nablaA_{\mu} \phi, \nablaA_{\mu}\nablaA_{\nu}\phi$, and 
\begin{align}
\GA{}^{\mu\nu\alpha\beta}&:=\frac{1}{4}\epsilon^{\mu\nu\rho\sigma} \epsilon^{\alpha\beta\gamma\delta} \RA{}_{\rho\sigma\gamma\delta}
\,, \\
\GA{}^{\mu\nu}&:=\GA{}^{\mu\alpha\nu}{}_{\alpha} \,.
\end{align}
are the dual Riemann tensor and the Einstein tensor, respectively.  

In the unitary gauge, $\nablaA_{\mu}\phi \propto n_{\mu}$, the relevant components of the non-minimal couplings \eqref{non_minimal} are
\begin{align}
\GA{}^{\mu\nu}n_{\mu}n_{\nu}\,, ~
\GA{}^{\mu\alpha\nu\beta}n_{\mu}n_{\nu}\gamma_{\alpha}^{\alpha'} \gamma_{\beta}^{\beta'} \,.
\end{align}
The $3+1$ decomposition shows that these terms do not have either $\pounds_n K_{\mu\nu}$ or $\pounds_n \kappa^{\alpha}{}_{\mu\nu}$ as explicitly shown in Appendix~\ref{appendix}.  Therefore, even if \eqref{non_minimal} are included, the Lagrangian still consists of
\begin{align}
t,N,\gamma_{\mu\nu},K_{\mu\nu},\kappa^{\alpha}{}_{\mu\nu},D_{\mu} \,,
\end{align}
and then the theory is free from the Ostrogradsky ghost.

We thus conclude that the Lagrangian,
\begin{align}
\mathcal{L}&=\frac{M_{\rm pl}^2}{2}\RA + f_2 \GA{}^{\mu\nu}\nablaA_{\mu}\phi \nablaA_{\nu} \phi 
\nn
&+ f_3 \GA{}^{\mu\alpha\nu\beta}\nablaA_{\mu}\phi \nablaA_{\nu} \phi \nablaA_{\alpha}\nablaA_{\beta}\phi +\mathcal{L}_{\phi}\,, \label{GF}
\end{align}
does not contain $\pounds_n A_*$ even after integrating out $\kappa$: the theory is a trivially U-degenerate theory.

\subsection{U-degenerate theories via conformal and disformal transformations}
We then consider theories whose Lagrangian explicitly contain $\pounds_n A_*$ after integrating out $\kappa$ but $\pounds_n A_*$ is degenerate and then free from the Ostrogradsky ghost.  The degeneracy of the kinetic matrix indicates that the kinetic matrix has zero eigenvalues and corresponding eigenvectors.  Hence, appropriately choosing the basic variables, the kinetic matrix may be block diagonalized and decomposed into the zero matrix and a non-degenerate matrix.  In the frame of the block diagonalization, the theory is trivially degenerate.  Conversely, we can generate a degenerate theory from a trivially degenerate theory via a change of the variables.  This was actually used to derive beyond Horndeski theories (degenerate theories) from the Horndeski theory (a trivially degenerate theory)~\cite{Zumalacarregui:2013pma} (see also~\cite{Bettoni:2013diz}).

We thus take field redefinitions to generate more general ghost-free theories from \eqref{GF}.  We note that the Lagrangian does not contain any derivatives of the metric when regarding that the metric and the connection are independent variables.  Therefore, the transformation of the metric with keeping the connection gives just an algebraic change in the metric-affine formalism.  Let us consider a conformal transformation,
\begin{align}
g_{\mu\nu} \rightarrow \bar{g}_{\mu\nu}=\Omega^2 g_{\mu\nu}\,,
\end{align}
where $\Omega$ is a function of $\phi$ and its $n$-th derivatives.
The theory \eqref{GF} is then transformed into
\begin{widetext}
\begin{align}
\sqrt{-\bar{g}} \mathcal{L}|_{g_{\mu\nu} \rightarrow \bar{g}_{\mu\nu}}=\sqrt{-g}\left[ \frac{M_{\rm pl}^2}{2} \Omega^2 
\RA + \bar{f}_2 \GA{}^{\mu\nu}\nablaA_{\mu}\phi \nablaA_{\nu} \phi + \Omega^{-2} \bar{f}_3 \GA{}^{\mu\alpha\nu\beta}\nablaA_{\mu}\phi \nablaA_{\nu} \phi \nablaA_{\alpha}\nablaA_{\beta}\phi +\Omega^4 \bar{\mathcal{L}}_{\phi}\right] \,, \label{GF2}
\end{align}
\end{widetext}
where the functions with a bar are the functions of the seed Lagrangian \eqref{GF}, $\bar{f}_2=\bar{f}_2(\bar{g},\phi,\nablaA_{\mu}\phi,\nablaA_{\mu}\nablaA_{\nu}\phi)=\bar{f}_2(\Omega^2 g,\phi,\nablaA_{\mu}\phi,\nablaA_{\mu}\nablaA_{\nu}\phi)$ and so on.

Although the change of the variables must not increase or decrease the number of degrees of freedom as long as the transformation is invertible and matter fields are not introduced, the ghost may appear when we consider an additional matter field $\psi$ (see e.g.,~\cite{Domenech:2015tca,Aoki:2018zcv}).  For instance, \eqref{GF2} with a minimal matter coupling
\begin{align}
\sqrt{-g}\left[ \frac{M_{\rm pl}^2}{2} \Omega^2 \RA + \cdots  +\mathcal{L}_{\rm m}(g, \Gamma, \psi) \right] \,,
\end{align}
is equivalent to \eqref{GF} with a non-minimal coupling
\begin{align}
\sqrt{-\bar{g}} \left[ \frac{M_{\rm pl}^2}{2} \RA  + \cdots + \Omega^{-4} \mathcal{L}_{\rm m}(\Omega^{-2}\bar{g},\Gamma, \psi) \right] \,.
\label{with_matter}
\end{align}
The simplest ``matter coupling'' is coupling to the cosmological constant
\begin{align}
\mathcal{L}_{\rm m} =-M_{\rm pl}^2 \Lambda \,, \label{cc}
\end{align}
which yields the term
\begin{align}
-\Omega^{-4}M_{\rm pl}^2 \Lambda
\end{align}
in the Lagrangian \eqref{with_matter}.  If $\Omega$ contains $\pounds_n A_*$, the theory \eqref{with_matter} with just a constant term \eqref{cc} is no longer ghost-free.  Hence, the conformal factor $\Omega$ is usually assumed to be a function up to the first derivative of $\phi$.  However, we have already shown that any functions with up to second-order covariant derivative of $\phi$ do not contain $\pounds_n A_*$ in the unitary gauge if the function is projective invariant.  The conformal factor can include the second-order derivatives when it does not contain the second-order time derivative.

We thus assume that $\Omega$ is projective invariant and is given by
\begin{align}
\Omega=\Omega(g,\phi,\nablaA_{\mu}\phi,\nablaA_{\mu}\nablaA_{\nu}\phi ) 
\end{align}
in order that the non-minimal couplings do not yield the Ostrogradsky ghost even after adding a matter field.  In particular, the term obtained from the cosmological constant \eqref{cc} is absorbed into the definition of $\mathcal{L}_{\phi}$.  
As a result, the conformal transformation generates a degenerate Lagrangian
\begin{align}
f_1
\RA + f_2 \GA{}^{\mu\nu}\nablaA_{\mu}\phi \nablaA_{\nu} \phi 
+ f_3 \GA{}^{\mu\alpha\nu\beta}\nablaA_{\mu}\phi \nablaA_{\nu} \phi \nablaA_{\alpha}\nablaA_{\beta}\phi +\mathcal{L}_{\phi}  \,, \label{GF4}
\end{align}
with four arbitrary functions $f_1,f_2,f_3,\mathcal{L}_{\phi}$ where 
\begin{align}
f_1=\frac{M_{\rm pl}^2}{2}\Omega^2 \,,~ f_2 =\bar{f}_2 \,, ~f_3=\bar{f}_3 \Omega^{-2} \,, ~ \mathcal{L}_{\phi}=\Omega^4 \bar{\mathcal{L}}_{\phi}
\,.  
\end{align}
As discussed in the beginning of this section, the non-minimal coupling to the Ricci curvature may yield $\pounds_n A_*$ after integrating out $\kappa$; however, since the Lagrangian is degenerate, \eqref{GF4} is free from the Ostrogradsky ghost as long as the unitary gauge can be imposed.

One may further consider a general transformation~\cite{Zumalacarregui:2013pma}
\begin{align}
g^{\mu\nu}\rightarrow \bar{g}^{\mu\nu}=\Omega^{-2}(g^{\mu\nu}+\Gamma^{\mu\nu}) \,,
\end{align}
where $\Gamma^{\mu\nu}$ is a symmetric and projective invariant tensor constructed by $g^{\mu\nu},\phi,\nablaA_{\mu}\phi,\nablaA_{\mu}\nablaA_{\nu}\phi$.  Furthermore, $g^{\mu\nu}$ is assumed to be non-degenerate.  We define the transformation of the inverse matrix for convenience.  Using the relation
\begin{align}
{\rm det} (\bar{g}_{\mu\nu})= \frac{\Omega^8}{{\rm det}(\delta^{\mu}_{\nu}+\Gamma^{\mu}_{\nu})} {\rm det } (g_{\mu\nu})\,,
\end{align}
one can straightforwardly calculate the transformed theory from \eqref{GF}.  Since the general expression is complicated due to the couplings to the Einstein tensor and the dual Riemann tensor, we just show the case of the disformal transformation~\cite{Bekenstein:1992pj}
\begin{align}
\Gamma^{\mu\nu}=\Gamma \nablaA{}^{\mu}\phi \nablaA{}^{\nu} \phi
\,,
\end{align}
whereas $\Gamma$ is a scalar function constructed by\\ $g^{\mu\nu},\phi,\nablaA_{\mu}\phi,\nablaA_{\mu}\nablaA_{\nu}\phi$.
Due to the fact that the Einstein tensor and the dual Riemann tensor couplings are given by the contraction of the Levi-Civita tensor, these couplings do not generate new kind of terms via the disformal transformation.  The only new coupling is obtained from the Ricci scalar.  The U-degenerate Lagrangian of the transformed theory is thus given by
\begin{align}
\mathcal{L}_{\rm UD}
&=
f_1
\RA + f_2 \GA{}^{\mu\nu}\nablaA_{\mu}\phi \nablaA_{\nu} \phi 
+ f_3 \GA{}^{\mu\alpha\nu\beta}\nablaA_{\mu}\phi \nablaA_{\nu} \phi \nablaA_{\alpha}\nablaA_{\beta}\phi 
\nn
&+ f_4 \RA{}_{\mu\nu}\nablaA{}^{\mu}\phi \nablaA{}^{\nu}\phi 
+\mathcal{L}_{\phi} \,, \label{GF3}
\end{align}
with
\begin{align}
f_1&= \frac{M_{\rm pl}^2\Omega^2}{2\sqrt{1+\Gamma X}} \,, ~
f_2=\bar{f}_2 \sqrt{1+\Gamma X} \,,~
\nn
f_3&=\Omega^{-2} \bar{f}_3 \sqrt{1+\Gamma X}\,, ~
f_4=\frac{M_{\rm pl}^2\Omega^2 \Gamma }{2\sqrt{1+\Gamma X}}  \,, ~
\nn
\mathcal{L}_{\phi}&=\Omega^4 \bar{\mathcal{L}}_{\phi} {\sqrt{1+\Gamma X}} \,,
\end{align}
where $X=g^{\mu\nu}\nablaA{}_{\mu}\phi \nablaA_{\nu} \phi$ and $\RA{}_{\mu\nu}=\RA{}^{\alpha}{}_{\mu\alpha\nu}$.
 This theory contains five arbitrary functions $f_1,f_2,f_3,f_4,\mathcal{L}_{\phi}$.  The regularity of the transformation imposes that $\Omega$ and $1+\Gamma X$ do not cross zero nor diverge which give restrictions $f_1, f_1+f_4 X \neq 0$.

Note that $\RA{}_{\mu\nu}\neq\RA{}_{\mu\alpha\nu}{}^{\alpha}$ in metric-affine geometry and the Einstein tensor is
\begin{align}
\GA{}_{\mu\nu}=\frac{1}{2}\left( \RA{}_{\mu\nu} + \RA{}_{\mu\alpha\nu}{}^{\alpha} -g_{\mu\nu}\RA \right)
\,.
\end{align}
Hence, the Ricci tensor coupling $\RA{}_{\mu\nu}\nablaA{}^{\mu}\phi \nablaA{}^{\nu}\phi $ cannot be absorbed into the the couplings to the Ricci scalar and the Einstein tensor via the redefinitions of the functions.

\section{Quadratic scalar-metric-affine theory}
\label{sec_quadratic}
We consider a concrete Lagrangian and show the kinetic structure after integrating out $\kappa$.  To explicitly solve the equation of the connection, we shall focus on up to the quadratic order of the connection.  Then, the most general quadratic order projective invariant Lagrangian consisting of the curvature, the scalar field $\phi$, and its derivatives is
\begin{widetext}
\begin{align}
\mathcal{L}_{\rm qPI}(g,\Gamma,\phi)&=f_1
\RA + f_2 \GA{}^{\mu\nu}\nablaA_{\mu}\phi \nablaA_{\nu} \phi + f_4 \RA{}_{\mu\nu}\nablaA{}^{\mu}\phi \nablaA{}^{\nu}\phi 
+F_2 +F_3 \mathcal{L}_3^{\rm gal \Gamma}
+F_4 \mathcal{L}_4^{\rm gal \Gamma}
\nn
&+C_1 \epsilon^{\mu\nu\rho\sigma}\epsilon^{\mu'\nu'\rho'}{}_{\sigma}\nablaA_{\mu}\phi \nablaA_{\mu'} \phi \nablaA{}_{\nu}\nablaA{}_{\nu'}\phi \nablaA{}_{[\rho}\nablaA{}_{\rho']}\phi
+C_2 ( \mathcal{L}_3^{\rm gal \Gamma})^2
\nn
&+C_3 (g^{\mu\beta}g^{\nu\delta}g^{\alpha\gamma}-g^{\mu\nu}g^{\alpha\gamma}g^{\beta\delta})\partial_{\mu}\phi \partial_{\nu} \phi \nablaA{}_{\alpha}\nablaA{}_{\beta}\phi \nablaA{}_{\gamma}\nablaA{}_{\delta} \phi\,,
\label{affine_action}
\end{align}
where 
\begin{align}
\mathcal{L}_3^{\rm gal \Gamma} &= \epsilon^{\alpha\beta\gamma\delta} \epsilon^{\alpha'\beta' }{}_{\gamma\delta} \nablaA{}_{\alpha} \phi \nablaA{}_{\alpha'} \phi  \nablaA{}_{\beta} \nablaA{}_{\beta'}\phi 
\,, \\
\mathcal{L}_4^{\rm gal \Gamma} &= \epsilon^{\alpha\beta\gamma\delta} \epsilon^{\alpha'\beta'\gamma'}{}_{\delta} \nablaA{}_{\alpha} \phi \nablaA{}_{\alpha'} \phi \nablaA{}_{\beta} \nablaA{}_{\beta'}\phi \nablaA{}_{\gamma} \nablaA{}_{\gamma'} \phi
\,, \label{galG4} 
\end{align}
are the projective invariant Galileon terms and $f_1,f_2,f_4,F_2,F_3,F_4,C_1,C_2,C_3$ are arbitrary functions of $\phi$ and $X:=(\partial \phi)^2$.  Note that, up to the quadratic order of the connection, the most general scalar-metric-affine theory is a class of the ghost-free scalar-metric-affine theory \eqref{GF3}.  We also notice that $C_1$ does not appear in the final expression and can set $C_1=0$ without loss of generality~\cite{Aoki:2018lwx}.

As shown in~\cite{Aoki:2018lwx}, the quadratic DHOST theory is obtained when $C_2=C_3=0$; in this case, all scalar self-interactions are given by the form of Galileon.  On the other hand, in the case $C_2,C_3\neq 0$, the Lagrangian is not a class of the quadratic DHOST.  However, we will explicitly show that \eqref{affine_action} indeed satisfies the degeneracy condition in the unitary gauge even when $C_2,C_3\neq 0$, and \eqref{affine_action} is equivalent to the quadratic U-degenerate theory.

After integrating out $\kappa$ from \eqref{affine_action}, we obtain the quadratic U-degenerate Lagrangian
\begin{align}
\mathcal{L}_{\rm qU}(g,\phi)&=f R(g)+P+Q_1 g^{\mu\nu} \phi_{\mu\nu} +Q_2\phi^{\mu} \phi_{\mu\nu}\phi^{\nu} 
+\left( \kappa_1 + \frac{f}{X} \right) L^{(2)}_1 +\left( \kappa_2 -\frac{f}{X}\right) L^{(2)}_2
\nn
&+ \left( \frac{2f}{X^2}-\frac{4f_X}{X}+2\sigma \kappa_1 +2 \left[ 3\sigma-\frac{1}{X} \right] \kappa_2 \right) L^{(2)}_3
+\left( \alpha +\frac{2f_X}{X}-\frac{2f}{X^2}-\frac{2\kappa_1}{X} \right) L^{(2)}_4
\nn
&+ \left( -\frac{\alpha}{X}+\frac{2f_X}{X^2}+\kappa_1 \left[ \frac{1}{X^2}+3\sigma^2 -\frac{2\sigma}{X} \right] +\kappa_2 \left[ 3\sigma -\frac{1}{X}\right]^2 \right) L^{(2)}_5
\,, \label{U-DHOST}
\end{align}
where
\begin{align}
L_1^{(2)}&=\phi_{\mu\nu}\phi^{\mu\nu}\,,~ L_2^{(2)}=(\phi^{\mu}{}_{\mu})^2 \,,~ 
L_3^{(2)}=\phi^{\mu}\phi^{\nu}\phi_{\mu\nu} \phi^{\rho}{}_{\rho} 
\,, ~
L_4^{(2)}=\phi_{\mu\nu}\phi^{\mu}\phi^{\nu\rho}\phi_{\rho}
\,, ~
L_5^{(2)}=(\phi^{\mu}\phi^{\nu}\phi_{\mu\nu})^2
\,.
\end{align}
with the notations $\phi_{\mu}=\nabla_{\mu}\phi,\phi_{\mu\nu}=\nabla_{\mu}\nabla_{\nu}\phi$.  The functions $f,P,Q_1,Q_2,\alpha,\kappa_1,\kappa_2,\sigma$ are given by
\begin{align}
f&=f_1-\frac{f_2 X}{2}
\,, \\
P&=F_2+\frac{3X(g_{\phi}-2F_3'X)^2}{8f g + 4X^2[C_3'+2(F_4'-6C_2' X)]}
\,, \\
Q_1&=-2f_{\phi}+\frac{2g(g_{\phi}-2F_3'X)}{2f g + X^2[C_3'+2(F_4'-6C_2' X)]}
\,, \\
Q_2&=\frac{2f_{\phi}}{X}-\frac{(g_{\phi}-2F_3'X)(2g-3g_X X)}{X[ 2f g + X^2 \{ C_3'+2(F_4'-6C_2' X) \} ]}
\,, \\
\alpha&=-\frac{f g_X (4g+g_X X)-4f_X g (g+2 g_X X)+2C_3' X(f^2-3f f_X X +4f_X^2 X^2)}{2g^2 X-C_3' f X^3}
\,, \\
\kappa_1&= -\frac{g^2}{f g X -(C_3'-F_4')X^3}
\,, \\
\kappa_2&= \frac{g^2 (2f g +X^2(2F_4'-4C_2' X-C_3'))}{X(fg - (C_3'-F_4')X^2)[2 f g +X^2\{ C_3'+2(F_4'-6C_2' X)\} ]}
\,, \\
\sigma&=\frac{g_X}{2g}
\end{align}
with
\begin{align}
g=f_1 (f_1+f_4 X)
\,,~F_3'=F_3 (f_1+f_4 X) \,, ~F_4'=F_4(f_1+f_4 X)^2\,,~ C_2'=C_2(f_1+f_4 X)^2\,, ~C_3'=C_3(f_1+f_4 X)^2\,.
\end{align}
In the unitary gauge, the Lagrangian is reduced to
\begin{align}
\mathcal{L}_{\rm qU}&= A_*^2 \hat{K}^{\mu\nu,\alpha\beta} \left( K_{\mu\nu} - \sigma \gamma_{\mu\nu} A_* \pounds_n A_* \right) \left( K_{\alpha\beta}-\sigma \gamma_{\alpha\beta} A_* \pounds_n A_* \right)
\nn
&+P-(Q_1-A_*^2 Q_2)\pounds_n A_* -A_* (2f_{\phi}+Q_1) K^{\mu}{}_{\mu}+f {}^3 \! R +(2f_X+A_*^2 \alpha)D_{\mu}A_* D^{\mu} A_*
\label{U-unitary}
\end{align}
where
\begin{align}
\hat{K}^{\mu\nu,\alpha\beta} = (\kappa_1 \gamma^{\mu(\alpha}\gamma^{\beta)\nu}+\kappa_2 \gamma^{\mu\nu}\gamma^{\alpha\beta} ) \,.  
\end{align}
\end{widetext}

We should emphasize that the cases $f_1=0$ and $f_1+f_4 X=0$ cannot be reduced to the Lagrangian \eqref{GF} via regular conformal/disformal transformations.  The dynamical degrees of freedom may not be three in these cases.  Indeed, the kinetic structure \eqref{U-unitary} shows that the case $g=0$ (which is the case when $f_1=0$ or $f_1+f_4 X=0$) leads to a totally degenerate theory where there are no dynamical degrees of freedom.

We can also see that the cases $\sigma=0$ ($g_X=0$) give trivially U-degenerate theories.  Because of the relation
\begin{align}
Q_1+X Q_2=g_X\cdot \frac{3  X(g_{\phi}-2 F_3' X)}{2 f g +X^2 [ C_3'+2(F_4'-6C_2' X) ] } \,,
\end{align} 
the unitary gauge Lagrangian \eqref{U-unitary} does not contain $\pounds_n A_*$ if $g_X=0$.  
This confirms the discussion done in Section~\ref{sec_non-minimal}: \eqref{GF} describes a trivially U-degenerate theory while \eqref{GF3} include U-degenerate theories.

\section{Concluding remarks}
\label{summary}
In this paper, we discussed the possibility that the Ostrogradsky ghost-free property of scalar-tensor theories are guaranteed by symmetry and show that the projective symmetry could be an important ingredient for ghost-free theories.  The projective transformation, which causes a shift of the affine connection, is defined in metric-affine geometry where the metric and the connection are independent.  We call scalar-tensor theories in the metric-affine geometry scalar-metric-affine theories since their independent variables are not only the scalar field and the metric but also the affine connection.  We consider two classes of the projective invariant scalar-metric-affine theories: theories without non-minimal coupling to the curvature and theories with.
In the former theories, we have shown that theories whose Lagrangian is constructed by up to second-order covariant derivatives are free from the Ostrogradsky ghost when the Lagrangian enjoys the projective symmetry and the unitary gauge can be imposed.  In the latter case, although we have not been able to conclude that general projective invariant scalar-metric-affine theories are ghost free, we found that a wide class of theories are free from the ghost.  The general ghost-free Lagrangian, which we found, contains five arbitrary scalar functions consisting of up to second-order derivatives of the scalar field\footnote{In Appendix~\ref{appendix}, we will discuss the existence of a more general ghost-free scalar-metric-affine theory.}.  

It would be worth emphasizing that the theories discussed in the present paper satisfy the degeneracy conditions at least in the unitary gauge while scalar-metric-affine theories constructed by Galileon type self-interactions satisfy the degeneracy conditions in any gauge (at least up to quartic Galileon)~\cite{Aoki:2018lwx}.  If one imposes to satisfy the degeneracy conditions in arbitrary gauge, another restriction on the Lagrangian in addition to the projective symmetry, such as the weakly broken Galileon symmetry~\cite{Pirtskhalava:2015nla,Santoni:2018rrx}, should be required.

We have also shown the explicit relation between the quadratic order scalar-metric-affine theories and the quadratic U-degenerate theories.  This reveals that the quadratic U-degenerate theories have the hidden projective symmetry which cannot be seen after integrating out the connection.  Furthermore, this explicit relation would be useful for phenomenology.  The metric-affine (Palatini) formalism of gravity have gained increasing attention in the highlight of constructing inflation models since the metric and the metric-affine (Palatini) formalisms compute different results and thus observations of inflation can potentially reveal what the fundamental variables of gravity are~\cite{Bauer:2008zj,Bauer:2010jg,Tamanini:2010uq,Rasanen:2017ivk,Tenkanen:2017jih,Racioppi:2017spw,Markkanen:2017tun,Jarv:2017azx,Racioppi:2018zoy,Carrilho:2018ffi,Enckell:2018kkc,Enckell:2018hmo,Almeida:2018oid,Antoniadis:2018ywb,Rasanen:2018fom,Kannike:2018zwn,Rasanen:2018ihz,Antoniadis:2018yfq,Shimada:2018lnm,Takahashi:2018brt,Jinno:2018jei,Iosifidis:2019jgi,Tenkanen:2019jiq,Rubio:2019ypq}.  In this context, theories which can have the Einstein frame have been mainly analyzed because the equation of the connection becomes simple in the Einstein frame.  Instead, we have already solved the equation of the connection to obtain \eqref{U-DHOST}.  One can analyze the phenomenology of the quadratic scalar-metric-affine theories by using the scalar-tensor theories without solving the equation of the connection.

In summary, we have highlighted the importance of the projective symmetry for ghost-free theories.  In particular, theories without non-minimal couplings to gravity are guaranteed to be free from the ghost by the projective symmetry.  It would be thus interesting to discuss whether the ghost-free structure is protected by the symmetry even in more general theories including complicated non-minimal couplings to the curvature as well as higher curvature terms.  Recently, the paper~\cite{BeltranJimenez:2019acz} showed the projective symmetry is required to make $f(\RA{}_{\mu\nu})$ theories ghost-free which also suggests the importance of the projective symmetry.  We may come back to the issue of clarifying further relations between the symmetry and the ghost-free property in generic theories in the future.

\section*{Acknowledgments}
We would like to thank Shinji Mukohyama, Kazuhumi Takahashi, Tommi Tenkanen and Masahide Yamaguchi for useful discussions and comments.
The work of K.A.  was supported in part by a Waseda University Grant for Special Research Projects (No.~2018S-128).



\appendix
\section{Hamiltonian analysis of scalar-metric-affine theories}
\label{appendix}
\subsection{$3+1$ decomposition of the curvature tensors}
Here, we summarize the $3+1$ decomposed metric-affine Riemann curvature tensor by using the distortion tensor.  We first define the $3+1$ decomposed components of the distortion tensor as follows:
\begin{align}
\kappa_* &:=\kappa_{\alpha\beta\gamma}n^{\alpha}n^{\beta}n^{\gamma} \,, \\
\hat{\kappa}^1_{\mu}&:=\kappa_{\alpha\beta\gamma}n^{\alpha} n^{\beta} \gamma^{\gamma}_{\mu} \,, \\
\hat{\kappa}^2_{\mu}&:=\kappa_{\alpha\beta\gamma}n^{\alpha} \gamma^{\beta}_{\mu} n^{\gamma} \,, \\
\hat{\kappa}^3_{\mu}&:=\kappa_{\alpha\beta\gamma}\gamma^{\alpha}_{\mu} n^{\beta}n^{\gamma} \,, \\
\hat{\kappa}^1_{\mu\nu}&:=\kappa_{\alpha\beta\gamma}n^{\alpha} \gamma^{\beta}_{\mu} \gamma^{\gamma}_{\nu}  \,, \\
\hat{\kappa}^2_{\mu\nu}&:=\kappa_{\alpha\beta\gamma}\gamma^{\alpha}_{\mu} n^{\beta} \gamma^{\gamma}_{\nu}  \,, \\
\hat{\kappa}^3_{\mu\nu}&:=\kappa_{\alpha\beta\gamma}\gamma^{\alpha}_{\mu} \gamma^{\beta}_{\nu} n^{\gamma} \,, \\
\hat{\kappa}_{\mu\nu\rho}&:=\kappa_{\alpha\beta\gamma}\gamma^{\alpha}_{\mu} \gamma^{\beta}_{\nu} \gamma^{\gamma}_{\rho} \,,
\end{align}
Under the projective transformation, $\kappa_{\mu\nu\rho}\rightarrow \kappa_{\mu\nu\rho}+g_{\mu\nu}U_{\rho}$, the $3+1$ decomposed components are transformed as
\begin{align}
\kappa_* &\rightarrow \kappa_* -U_* 
\,, ~\hat{\kappa}^3_{\mu\nu}\rightarrow \hat{\kappa}^3_{\mu\nu}+\gamma_{\mu\nu}U_*
\,, 
\nn
\hat{\kappa}^1_{\mu} &\rightarrow \hat{\kappa}^1_{\mu}-\hat{U}_{\mu}
\,, ~\hat{\kappa}_{\mu\nu\rho}\rightarrow \hat{\kappa}_{\mu\nu\rho}+\gamma_{\mu\nu} \hat{U}_{\rho}
\end{align}
and other components are unchanged where
\begin{align}
U_*:=U_{\alpha}n^{\alpha}\,, ~ \hat{U}_{\mu}:=U_{\alpha}\gamma^{\alpha}_{\mu} \,.
\end{align}
To decompose the Riemann curvature, it is useful to define the variables,
\begin{align}
\KA{}^1_{\mu\nu}&:=(\nablaA_{\beta}n_{\alpha})\gamma^{\alpha}_{\mu}\gamma^{\beta}_{\nu}
=K_{\mu\nu}-\hat{\kappa}^1_{\mu\nu}
\,, \\
\KA{}^2_{\mu\nu}&:=(\nablaA_{\beta}n^{\alpha})\gamma_{\alpha\mu}\gamma^{\beta}_{\nu}
=K_{\mu\nu}+\hat{\kappa}^2_{\mu\nu}
\,,
\end{align}
which could be considered as the extrinsic curvature of metric-affine geometry.
Then, all independent components of the $3+1$ decomposed curvature are 
\newpage
\begin{widetext}
\begin{align}
\RA{}_{\alpha\beta\gamma\delta}n^{\alpha}n^{\beta}n^{\gamma}\gamma^{\delta}_{\rho}
&=\pounds_n \hat{\kappa}^1_{\rho}-D_{\rho}\kappa_* -a_{\rho} \kappa_* +a^{\gamma}(\KA{}^1_{\gamma\rho} -\KA{}^2_{\gamma\rho})
+\KA{}^{1\gamma}{}_{\rho}\hat{\kappa}^3_{\gamma}+\KA{}^{2\gamma}{}_{\rho}\hat{\kappa}^2_{\gamma}
\,,
\\
\RA{}_{\alpha\beta\gamma\delta}n^{\alpha}\gamma^{\beta}_{\mu}n^{\gamma}\gamma^{\delta}_{\rho}
&=-\pounds_n \KA{}^1_{\mu\rho}-D_{\rho}\hat{\kappa}^2_{\mu}+\KA{}^1_{\alpha\rho}K^{\alpha}{}_{\mu}-a^{\alpha}\hat{\kappa}_{\alpha\mu\rho}-a_{\mu}\hat{\kappa}^1_{\rho}-a_{\rho}\hat{\kappa}^2_{\mu}+\KA{}^1_{\mu\rho}\kappa_*+\KA{}^{1\alpha}{}_{\rho}\hat{\kappa}^3_{\alpha \mu}+\hat{\kappa}^1_{\rho}\hat{\kappa}^2_{\mu} +\hat{\kappa}^{\alpha}{}_{\mu\rho}\hat{\kappa}^2_{\alpha}
\\
\RA{}_{\alpha\beta\gamma\delta}\gamma^{\alpha}_{\mu} n^{\beta}n^{\gamma}\gamma^{\delta}_{\rho}
&=\pounds_n \KA{}^2_{\mu\rho}-D_{\rho}\hat{\kappa}^3_{\mu}-\KA{}^2_{\alpha\rho}K^{\alpha}{}_{\mu}-a^{\alpha}\hat{\kappa}_{\mu\alpha\rho}-a_{\mu}\hat{\kappa}^1_{\rho}-a_{\rho}\hat{\kappa}^3_{\mu}+\KA{}^2_{\mu\rho}\kappa_*+\KA{}^{2\alpha}{}_{\rho}\hat{\kappa}^3_{\mu\alpha}-\hat{\kappa}^1_{\rho}\hat{\kappa}^3_{\mu} -\hat{\kappa}_{\mu}{}^{\alpha}{}_{\rho}\hat{\kappa}^3_{\alpha}
\,,
\\
\RA{}_{\alpha\beta\gamma\delta}\gamma^{\alpha}_{\mu}\gamma^{\beta}_{\nu}n^{\gamma}\gamma^{\delta}_{\rho}
&=\pounds_n \hat{\kappa}_{\mu\nu\rho}-D_{\rho}\hat{\kappa}^3_{\mu\nu}-2D_{[\mu}K_{\nu]\rho}-a_{\mu}(K_{\nu\rho}-\KA{}^1_{\nu\rho})+
a_{\nu}(K_{\mu\rho}-\KA{}^2_{\mu\rho})-a_{\rho}\hat{\kappa}^3_{\mu\nu}
\nn
&+(\hat{\kappa}^{3}_{\mu}{}^{\alpha}-K_{\mu}{}^{\alpha})\hat{\kappa}_{\alpha\nu\rho}-(\hat{\kappa}^{3\alpha}{}_{\nu}+K^{\alpha}{}_{\nu})\hat{\kappa}_{\mu\alpha\rho}
+\KA{}^1_{\nu\rho}\hat{\kappa}^3_{\mu}+\KA{}^2_{\mu\rho}\hat{\kappa}^2_{\nu}\,,
\\
\RA{}_{\alpha\beta\gamma\delta}n^{\alpha}n^{\beta} \gamma^{\gamma}_{\rho} \gamma^{\delta}_{\sigma}
&=2D_{[\rho}\hat{\kappa}^1_{\sigma]}-2\KA{}^{1\gamma}{}_{[\rho}\KA{}^2_{|\gamma|\sigma]}
\,,
\\
\RA{}_{\alpha\beta\gamma\delta}n^{\alpha}\gamma^{\beta}_{\mu} \gamma^{\gamma}_{\rho} \gamma^{\delta}_{\sigma}
&=-2D_{[\rho}\KA{}^1_{|\mu|\sigma]}-2\KA{}^{1}{}_{\mu[\rho}\hat{\kappa}^1_{\sigma]}-2\KA{}^{1\alpha}{}_{[\rho}\hat{\kappa}_{|\alpha\mu|\sigma]}
\,,
\\
\RA{}_{\alpha\beta\gamma\delta}\gamma^{\alpha}_{\mu} n^{\beta} \gamma^{\gamma}_{\rho} \gamma^{\delta}_{\sigma}
&=2D_{[\rho}\KA{}^2_{|\mu|\sigma]}-2\KA{}^2_{\mu[\rho}\hat{\kappa}^1_{\sigma]}-2\KA{}^{2\alpha}{}_{[\rho}\hat{\kappa}_{|\mu\alpha|\sigma]}
\,,
\\
\RA{}_{\alpha\beta\gamma\delta}\gamma^{\alpha}_{\mu}\gamma^{\beta}_{\nu} \gamma^{\gamma}_{\rho} \gamma^{\delta}_{\sigma}
&=\RAcal{}_{\mu\nu\rho\sigma}+2\KA{}^1_{\nu[\sigma}\KA{}^2_{|\mu|\rho]}
\,.
\end{align}
where 
\begin{align}
\RAcal{}_{\mu\nu\rho\sigma}:=\mathcal{R}_{\mu\nu\rho\sigma}(\gamma)+2D_{[\rho}\hat{\kappa}_{\mu\nu|\sigma]}
+2\hat{\kappa}_{\mu}{}^{\alpha}{}_{[\rho}\hat{\kappa}_{\alpha\nu|\sigma]}
\end{align}
and $\mathcal{R}_{\mu\nu\rho\sigma}(\gamma)$ is the spatial curvature constructed by the spatial metric $\gamma_{\mu\nu}$.
We then obtain
\begin{align}
\GA{}^{\mu\nu}n_{\mu}n_{\nu}&=\frac{1}{2}\left(\RAcal{}^{\mu\nu}{}_{\mu\nu}+\KA{}^{1\mu}{}_{\mu}\KA{}^{2\mu}{}_{\mu}-\KA{}^{1\mu\nu}\KA{}^2_{\nu\mu} \right)
\,,
\\
\GA{}^{\mu\alpha\nu\beta}n_{\mu}n_{\nu}\gamma_{\alpha}^{\alpha'} \gamma_{\beta}^{\beta'}
&=\frac{1}{2}\Biggl[ \KA{}^{1\mu\alpha'}\KA{}^{2\beta'}{}_{\mu} + \KA{}^{1\beta'\mu}\KA{}^2{}_{\mu}{}^{\alpha'}
-\KA{}^{1\mu}{}_{\mu}\KA{}^{2\beta'\alpha'} - \KA{}^{1\beta'\alpha'}\KA{}^{2\mu}{}_{\mu}
+\left(\KA{}^{1\mu}{}_{\mu}\KA{}^{2\mu}{}_{\mu}-\KA{}^{1\mu\nu}\KA{}^2_{\nu\mu}\right)\gamma^{\alpha'\beta'}
\nn
&\qquad \quad -\RAcal{}_{\mu}{}^{\beta'\mu\alpha'} - \RAcal{}^{\beta'\mu\alpha'}{}_{\mu}+\RAcal{}^{\mu\nu}{}_{\mu\nu}\gamma^{\alpha'\beta'} \Biggl] \,,
\end{align}
which show that the couplings \eqref{non_minimal} do not have time derivatives of the distortion tensor in the unitary gauge.

\subsection{Counting the number of degrees of freedom}
In the unitary gauge, the second-order derivative of the scalar field is
\begin{align}
\nablaA{}_{\mu}\nablaA{}_{\nu}\phi &=n_{\mu}n_{\nu}(\pounds_n A_* +A_* \kappa_*) -n_{\mu}(D_{\nu}A_*+A_* \hat{\kappa}^2_{\nu})
-n_{\nu}(D_{\mu}A_*+A_* \hat{\kappa}^1_{\mu})-A_* \KA{}^1_{\nu\mu} \,.
\end{align}
Recall that $\kappa_*$ and $\hat{\kappa}^1_{\mu}$ are projective modes which never appear in the projective invariant functions.  We define a new variable
\begin{align}
V_{\mu}:=\hat{\kappa}^2_{\mu}-D_{\mu}N/N\,.
\end{align}
Then, the second-order derivative is expressed by
\begin{align}
\nablaA{}_{\mu}\nablaA{}_{\nu}\phi = -A_* n_{\mu} V_{\nu}-A_* \KA{}^1_{\nu\mu} +{\rm projective~modes}\,,
\end{align}
which indicates that any projective invariant functions including up to the second-order derivatives are algebraic functions of $t, N, \gamma^{\mu\nu}, V_{\mu}, \KA{}^1_{\mu\nu}$ in the unitary gauge.

Here, we consider the Lagrangian
\begin{align}
\mathcal{L}=F^{\mu\nu\rho\sigma}\RA{}_{\mu\nu\rho\sigma}+\mathcal{L}_{\phi}
\,, \label{gLag}
\end{align}
where $F^{\mu\nu\rho\sigma}$ and $\mathcal{L}_{\phi}$ are constructed by $g^{\mu\nu},\phi,\nablaA_{\mu}\phi,\nablaA_{\mu}\nablaA_{\nu}\phi$.  Due to the antisymmetric property of the last two indices of the curvature tensor, the tensor $F^{\mu\nu\rho\sigma}$ is assumed to be $F^{\mu\nu(\rho\sigma)}=0$ without loss of generality.  For simplicity, we assume $F^{\mu\nu\rho\sigma}$ is given the form
\begin{align}
F^{\mu\nu\rho\sigma}=\hat{F}^{\mu\nu\rho\sigma}+ F_1 \gamma^{\nu[\rho}n^{\sigma]}n^{\mu} - F_2 \gamma^{\mu[\rho} n^{\sigma]}n^{\nu} 
+\hat{F}^{\alpha}{}_{\alpha}{}^{\rho\sigma} n^{\mu}n^{\nu} \,,
\label{exF}
\end{align}
with $F_1,F_2 \neq 0$ and $\hat{F}^{\mu\nu(\rho\sigma)}=0$ in the unitary gauge.  Since the projective transformation of the curvature tensor is
\begin{align}
\RA{}_{\mu\nu\rho\sigma} \rightarrow \RA{}_{\mu\nu\rho\sigma}+2g_{\mu\nu}\partial_{[\mu}U_{\nu]}
\end{align}
the property $g_{\mu\nu}F^{\mu\nu\rho\sigma}=0$ leads to that the projective invariance of the Lagrangian is guaranteed when $F^{\mu\nu\rho\sigma}$ and $\mathcal{L}_{\phi}$ are projective invariant.  The functions $F_1,F_2,\hat{F}^{\mu\nu\rho\sigma}$ and $\mathcal{L}_{\phi}$ are thus functions of $t, N, \gamma^{\mu\nu}, V_{\mu}, \KA{}^1_{\mu\nu}$ in the unitary gauge.

In the unitary gauge, \eqref{GF3} is given by 
\begin{align}
\mathcal{L}_{\rm UD}=\left[ \hat{F}_{\rm UD}^{\mu\nu\rho\sigma} +f_1 \gamma^{\nu[\rho}n^{\sigma]}n^{\mu} -(f_1-A_*^2 f_4)\gamma^{\mu[\rho}n^{\sigma]}n^{\nu} \right] \RA{}_{\mu\nu\rho\sigma}+\mathcal{L}_{\phi}
\end{align}
with
\begin{align}
\hat{F}_{\rm UD}^{\mu\nu\rho\sigma}=
 \frac{1}{2}(2f_1 + A_*^2 f_2 - A_*^3 f_3 \KA{}^{1\alpha}{}_{\alpha} ) \gamma^{\mu[\rho}\gamma^{\sigma]\nu} +\frac{1}{2}A_*^3 f_3 \KA{}^{1\mu[\rho}\gamma^{\sigma]\nu}-\frac{1}{2}A_*^3 f_3 \KA{}^{1\nu[\rho}\gamma^{\sigma]\mu}
\end{align}
Therefore, \eqref{GF3} is a subclass of \eqref{gLag} with \eqref{exF}.  We will show the ghost-freeness of the more general theory \eqref{gLag} instead of \eqref{GF3} itself.

After taking integration by part, the unitary gauge Lagrangian is given by
\begin{align}
N\mathcal{L}=N\Big[
&F_1 \gamma^{\mu\nu}\pounds_n \KA{}^1_{\mu\nu}
+F_2 \gamma^{\mu\nu}\pounds_n \KA{}^2_{\mu\nu}
-(F_1 \KA{}^{1\mu\nu} + F_2 \KA{}^{2\mu\nu})K_{\mu\nu}
\nn
&
+\hat{F}^{\mu\nu\rho\sigma}  (\mathcal{R}_{\mu\nu\rho\sigma} + 2D_{\rho}\kappa^{\rm PI}_{\mu\nu\sigma}
+2\kappa^{\rm PI}_{\mu}{}^{\alpha}{}_{\rho} \kappa^{\rm PI}_{\alpha\nu\sigma} +2 \KA{}^1_{\nu\sigma} \KA{}^2_{\mu\rho} )
-2\hat{F}_{\sigma}{}^{\sigma \mu\nu} \KA{}^1_{\rho\mu}\KA{}^{2 \rho}{}_{\nu}
\nn
&
-F_1 V_{\mu} \kappa^{ {\rm PI} \mu\nu}{}_{\nu}-V_{\mu}D^{\mu}F_1 +D^2 F_1 +\kappa^{{\rm PI} \mu\nu}{}_{\mu}D_{\nu}F_2 +F_2 D_{\mu} \kappa^{{\rm PI} \nu\mu}{}_{\nu}
\nn
& -(F_1 \KA{}^1_{\mu\nu} -F_2 \KA{}^2_{\nu\mu})\kappa^{{\rm PI} \mu\nu} +(D^{\nu}F_2 -F_2 \kappa^{{\rm PI} \mu\nu}{}_{\mu} )\hat{\kappa}^3_{\nu}
+\mathcal{L}_{\phi}
\Big]
\end{align}
where
\begin{align}
\kappa^{\rm PI}_{\mu\nu\rho}:=\hat{\kappa}_{\mu\nu\rho}+\gamma_{\mu\nu}\hat{\kappa}^1_{\rho}
\,, \quad
\kappa^{\rm PI}_{\mu\nu}:=\hat{\kappa}^3_{\mu\nu}+\gamma_{\mu\nu}\kappa_*
\end{align}
are projective invariant variables.  Since $F_1,F_2,\hat{F}^{\mu\nu\rho\sigma}, \mathcal{L}_{\phi}$ depends only on $t, N, \gamma^{\mu\nu}, V_{\mu}, \KA{}^1_{\mu\nu}$, the Lagrangian is linear in $\kappa^{\rm PI}_{\alpha\beta}$ and $\hat{\kappa}^3_{\mu}$ which act as the Lagrangian multipliers.

Time derivatives of the variables are given by the Lie derivatives with respect to the vector $t^{\mu}$ defined by
\begin{align}
t^{\mu}=N n^{\mu}+N^{\mu}
\end{align}
where $N$ and $N^{\mu}$ are the lapse and the shift vectors.  Introducing the 116 canonical variables 
\begin{align*}
(N,\pi_N),~
(N^{\mu},\pi_{\mu}),~
(\gamma_{\mu\nu},\pi^{\mu\nu}), ~
(\KA{}^1_{\mu\nu}, \Pi_1^{\mu\nu}), ~
(\KA{}^2_{\mu\nu}, \Pi_2^{\mu\nu}), ~
(\kappa^{\rm PI}_{\mu\nu\rho}, \Pi_{\rm PI}^{\mu\nu\rho}), ~
(V_{\mu}, \Pi^{\mu})\,,
\end{align*}
we have the 70 primary constraints given by
\begin{align}
\pi_N \approx 0 \,, \quad 
\pi_{\mu} \approx 0 \,, \quad
\Pi_{\rm PI}^{\mu\nu\rho}  \approx 0 \,, \quad
\Pi^{\alpha}  \approx 0 \,, \nn
\Pi_1^{\mu\nu}- \sqrt{\gamma} F_1 \gamma^{\mu\nu} \approx 0 \,, \quad
\Pi_2^{\mu\nu}-\sqrt{\gamma} F_2 \gamma^{\mu\nu} \approx 0 \,, \nn
\pi^{\mu\nu} + \frac{\sqrt{\gamma}}{2} (F_1 \KA{}^{1 (\mu\nu)} +F_2 \KA{}^{2 (\mu\nu)} ) \approx 0\,,
\label{p1}
\end{align}
and
\begin{align}
F_1 \KA{}^1_{\mu\nu} -F_2 \KA{}^2_{\nu\mu} &\approx 0 
\,, ~
D^{\nu}F_2 -F_2 \kappa^{{\rm PI} \mu\nu}{}_{\mu}   \approx 0\,,
\label{p2}
\end{align}
where the last two constraints are obtained because of the existence of the Lagrangian multipliers $\kappa^{\rm PI}_{\mu\nu}$ and $\hat{\kappa}^3_{\mu}$.  Due to the constraints $\Pi_1^{\mu\nu}- \sqrt{\gamma} F_1 \gamma^{\mu\nu} \approx 0$ and $\Pi_2^{\mu\nu}-\sqrt{\gamma} F_2 \gamma^{\mu\nu} \approx 0$, the functions $F_1$ and $F_2$ can be replaced with $\tilde{\Pi}_1:=\gamma_{\mu\nu} \Pi_1^{\mu\nu}/\sqrt{\gamma}$ and $\tilde{\Pi}_2:=\gamma_{\mu\nu} \Pi_2^{\mu\nu}/\sqrt{\gamma}$ by redefining the Lagrangian multipliers.  The total Hamiltonian is thus
\begin{align}
H_{\rm tot}=\int d^3 x \left( \mathcal{H}_V +\lambda^{\mu}\pi_{\mu}+N^{\mu} \mathcal{H}_{\mu} +\lambda_I \Pri{}^I \right)\,,
\end{align}
with
\begin{align}
\mathcal{H}_V=-N \sqrt{\gamma}
\Big[
&
\hat{F}^{\mu\nu\rho\sigma}  (\mathcal{R}_{\mu\nu\rho\sigma} + 2D_{\rho}\kappa^{\rm PI}_{\mu\nu\sigma}
+2\kappa^{\rm PI}_{\mu}{}^{\alpha}{}_{\rho} \kappa^{\rm PI}_{\alpha\nu\sigma} +2 \KA{}^1_{\nu\sigma} \KA{}^2_{\mu\rho} )
-2\hat{F}_{\sigma}{}^{\sigma \mu\nu} \KA{}^1_{\rho\mu}\KA{}^{2 \rho}{}_{\nu}
\nn
&
-\frac{1}{3} \tilde{\Pi}_1 V_{\mu} \kappa^{ {\rm PI} \mu\nu}{}_{\nu}- \frac{1}{3} V_{\mu}D^{\mu}\tilde{\Pi}_1 +\frac{1}{3} D^2 \tilde{\Pi}_1 +\frac{1}{3} \kappa^{{\rm PI} \mu\nu}{}_{\mu}D_{\nu}\tilde{\Pi}_2 +\frac{1}{3} \tilde{\Pi}_2 D_{\mu} \kappa^{{\rm PI} \nu\mu}{}_{\nu} +\mathcal{L}_{\phi} \Big]\,,
\end{align}
and
\begin{align}
\mathcal{H}_{\mu}(x)=\frac{\delta}{\delta N^{\mu}(x)} \int d^3y  P^A(y) \pounds_{N} Q_A(y) \,, \label{momentum}
\end{align}
where $\pounds_{N}$ is the Lie derivative with respect to the shift, and $Q_A$ and $P^A$ are the sets of the canonical variables, $Q_A=\{N,\gamma_{\mu\nu},\KA{}^1_{\mu\nu},\KA{}^2_{\mu\nu},\kappa^{\rm PI}_{\mu\nu\rho},V_{\mu}\}$ and $P^A=\{\pi_N,\pi^{\mu\nu},\Pi_1^{\mu\nu},\Pi_2^{\mu\nu},\Pi_{\rm PI}^{\mu\nu\rho},\Pi^{\mu}\}$.  $\Pri{}^I$ and $\lambda_I$ are the sets of the 67 primary constraints and the associated Lagrangian multipliers explicitly given by
\begin{align}
\lambda_I \Pri{}^I&=
\lambda_N \pi_N +\lambda^{\rm PI}_{\mu\nu\rho} \Pi_{\rm PI}^{\mu\nu\rho}+ \lambda^V_{\mu} \Pi^{\mu}
+\lambda^1_{\mu\nu}(\Pi_1^{\mu\nu}- \sqrt{\gamma} F_1 \gamma^{\mu\nu} )
+\lambda^2_{\mu\nu}(\Pi_2^{\mu\nu}-\sqrt{\gamma} F_2 \gamma^{\mu\nu})
\nn
&+\lambda_{\mu\nu}\left[ \pi^{\mu\nu} + \frac{\sqrt{\gamma}}{6} (\tilde{\Pi}_1 \KA{}^{1 (\mu\nu)} +\tilde{\Pi}_2 \KA{}^{2 (\mu\nu)} ) \right]
+\frac{1}{3} \sqrt{\gamma} \lambda_{\kappa}^{\mu\nu}( \tilde{\Pi}_1 \KA{}^1_{\mu\nu} -\tilde{\Pi}_2 \KA{}^2_{\nu\mu} )
+\frac{1}{3} \sqrt{\gamma} \lambda_{\kappa}^{\nu}( D_{\nu}\tilde{\Pi}_{2} -\tilde{\Pi}_{2} \kappa^{\rm PI}_{\mu\nu}{}^{\mu} ) \,,
\end{align}
where we have introduced
\begin{align}
\lambda_{\kappa}^{\mu\nu}=N \kappa^{{\rm PI} \mu\nu} \,, \quad
\lambda_{\kappa}^{\mu}=-N \hat{\kappa}^{3\mu} \,,
\end{align}
in order to emphasize that these two variables are just the Lagrangian multipliers.

The time preservation of $\pi_{\mu}\approx 0$ leads to the momentum constraint $\mathcal{H}_{\mu} \approx 0$ which are surely first class due to the existence of the spatial diffeomorphism invariance.  The time preservations of the other constraints yield
\begin{align}
\frac{d}{dt} \Pri{}^I(t, x) 
&\approx \int d^3 y \lambda_J(y) \mathcal{M}^{IJ}(t,x,y) + \{ \Pri{}^I (t,x), \int d^3 y \mathcal{H}_V(t, y) \} +\frac{\partial}{\partial t} \Pri{}^I(t,x) \approx 0 \,.
\end{align} 
where
\begin{align}
\mathcal{M}^{IJ}:= \{ \Phi^I(t,x) ,   \Phi^J(t, y) \}
\end{align}
If $\mathcal{M}^{IJ}$ has zero eigenvalues, some of the Lagrangian multipliers are undetermined and then the preservations of the primary constraints generate the secondary constraints when $\frac{d}{dt} \Pri{}^I \approx 0$ are not trivially satisfied.

To explicitly discuss how many secondary constraints are obtained from $\frac{d}{dt} \Pri {}^I \approx 0$, we first consider a simple case
\begin{align}
\hat{F}^{\mu\nu\rho\sigma} = \hat{F}  \gamma^{\mu [\rho} \gamma^{\nu| \sigma]} 
\end{align}
with $F_1,F_2,\hat{F} ={\rm constant}$.  In this case, the 31 components of the Lagrangian multipliers,
\begin{align}
\lambda_N,\quad
 \lambda^{\mu}_{\kappa},  \quad
F_1 \lambda^1_{[\mu\nu]} - F_2 \lambda^2_{[\mu\nu]}, \quad
\lambda^{\rm PI}_{\mu\nu\rho}-\frac{1}{3} \gamma_{\mu\rho} \lambda^{\rm PI}_{\sigma\nu}{}^{\sigma}
\end{align}
are undetermined and there exist the following 31 secondary constraints:
\begin{align}
\frac{d}{dt} \pi_N & \approx \mathcal{L}_{\phi}+N\frac{\partial \mathcal{L}_{\phi}}{\partial N}
-F_1 V^{\alpha} \kappa^{\rm PI}_{\alpha\beta}{}^{\beta}
\nn
&+\hat{F} \big( \KA{}^{1\alpha}{}_{\alpha} \KA{}^{2\beta}{}_{\beta} - \KA{}^{1\alpha\beta}\KA{}^2_{\beta\alpha}+\mathcal{R}(\gamma)-\kappa^{\rm PI}{}^{\alpha\beta\gamma} \kappa^{\rm PI}_{\beta\gamma\alpha} + D_{\alpha} \kappa{}^{\rm PI}{}^{\alpha\beta}{}_{\beta} \big)
\approx 0 
\,, \label{second1}
\\
\frac{d}{dt} \Pi^{\mu} &\approx
 N\left( \frac{\partial \mathcal{L}_{\phi}}{\partial V_{\mu}} -F_1 \kappa^{\rm PI}{}^{\mu\alpha}{}_{\alpha} \right) \approx 0 
 \,, \\
 \frac{1}{F_1} \frac{d}{dt} \Pi_1^{[\mu\nu]} -  \frac{1}{F_2} \frac{d}{dt} \Pi_2^{[\mu\nu]}
 &\approx -N\left( \frac{2\hat{F}}{F_2} \KA{}^{1[\mu\nu]} - \frac{1}{F_1} \frac{\partial \mathcal{L}_{\phi}}{\partial \KA{}^1_{[\mu\nu]}} \right) \approx 0
 \,, \\ 
\frac{d}{dt} \left( \Pi^{\mu\nu\rho} -\frac{1}{3}\gamma^{\mu\rho} \Pi^{\alpha\nu}{}_{\alpha} \right)
& \approx  -N F_1 V^{\mu}\gamma^{\nu\rho}- N \hat{F} \kappa{}^{\rm PI}{}^{\nu\rho\mu} -N \hat{F} \kappa^{\rm PI}{}^{\rho \mu \nu}
 -\hat{F} \gamma^{\nu\rho} D^{\mu} N
\nn
&+\frac{1}{3} \gamma^{\mu\rho} \left( N F_1 V^{\nu} +N\hat{F} \kappa{}^{\rm PI}_{\alpha}{}^{\alpha\nu} +N\hat{F} \kappa{}^{\rm PI}{}^{\nu\alpha}{}_{\alpha} + \hat{F} D^{\nu} N \right) 
\approx 0 \,.  \label{second4}
\end{align}
whose set is denoted by $\Psi{}^ i \approx 0$.
As a result, the number of phase space degrees of freedom is
\begin{align}
\leq 116-
\underbrace{6\times 2}_{\pi_{\mu} \approx 0, \mathcal{H}_{\mu} \approx 0 }
 - 
\underbrace{67}_{\Pri{}^I \approx 0} 
- 
\underbrace{31}_{\Sec{}^i \approx 0} = 6 \,.
\end{align}
which is sufficient to conclude that the Lagrangian is free from the Ostrogradsky ghost mode\footnote{There could be a first class constraint in $\Pri{}^I \approx 0 , \Sec{}^i \approx 0 $ or a tertiary constraint from $\frac{d}{dt} \Sec \approx 0$ depending on the Lagrangian which give further reductions of the number of degrees of freedom.  The resultant theory would be a class of the minimally modified gravity theories~\cite{Lin:2017oow,Aoki:2018zcv} or the cuscuton theories~\cite
{Afshordi:2006ad,Gomes:2017tzd,Iyonaga:2018vnu}.  However, we do not discuss whether such additional reductions exist and only focus on the fact the the Lagrangian is free from the Ostrogradsky ghost.}.

For generic functions of $F_1,F_2,\hat{F}^{\mu\nu\rho\sigma}$, the expressions become quite complicated.  Nonetheless, the structure of the constraints are not so changed.  Indeed, we confirm that the Lagrangian multipliers
\begin{align}
\lambda_N,\quad
 \lambda^{\mu}_{\kappa},  \quad
F_1 \lambda^1_{[\mu\nu]} - F_2 \lambda^2_{[\mu\nu]}, \quad
\lambda^{\rm PI}_{\mu\nu\rho}-\frac{1}{3} \gamma_{\mu\rho} \lambda^{\rm PI}_{\sigma\nu}{}^{\sigma}
\end{align}
are undetermined while others are determined even for the generic cases.  Eqs.~\eqref{second1}-\eqref{second4} become
\begin{align}
\frac{d}{dt}\pi_N &\approx \frac{\partial F_1}{\partial N} \lambda^{1\alpha}{}_{\alpha} + \frac{\partial F_2}{\partial N} \lambda^{2\alpha}{}_{\alpha} 
+\cdots
\approx 0
\,, \label{second1g} \\
\frac{d}{dt} \Pi^{\mu} &\approx \frac{\partial F_1}{\partial V_{\mu}} \lambda^{1\alpha}{}_{\alpha}  + \frac{\partial F_2}{\partial V_{\mu}} \lambda^{2\alpha}{}_{\alpha}  +\cdots 
\approx 0
\,, \\
\frac{1}{F_1} \frac{d}{dt} \Pi_1^{[\mu\nu]} -  \frac{1}{F_2} \frac{d}{dt} \Pi_2^{[\mu\nu]}
&\approx 
\frac{1}{F_1} \frac{\partial F_1}{\partial \KA{}^1_{[\mu\nu]}} \lambda^{1\alpha}{}_{\alpha}  + \frac{1}{F_1} \frac{\partial F_2}{\partial \KA{}^1_{[\mu\nu]}} \lambda^{2\alpha}{}_{\alpha}  
+\cdots
\approx 0
\,, \\
\frac{d}{dt} \left( \Pi^{\mu\nu\rho} -\frac{1}{3}\gamma^{\mu\rho} \Pi^{\alpha\nu}{}_{\alpha} \right)
&\approx \cdots 
\approx 0 \label{second4g}
\end{align}
\end{widetext}
where $\cdots$ stands for the terms not including the Lagrangian multipliers.  Since $\lambda^{1\alpha}{}_{\alpha}$ and $\lambda^{2\alpha}{}_{\alpha}$ are determined by the other components of $\frac{d}{dt} \Pri{}^I \approx 0$, Eqs.~\eqref{second1g}-\eqref{second4g} also give the 31 secondary constraints as with the previous example.  Therefore, the Lagrangian \eqref{gLag} with \eqref{exF} has at most 6 degree of freedom in the phase space and then free from the Ostrogradsky ghost.

\bibliography{ref}
\bibliographystyle{JHEP}

\end{document}